**Into the fission valley of "magic nucleus" Polonium**

Tilak Kumar Ghosh

Variable Energy Cyclotron Centre, 1/AF, Bidhannagar, Kolkata 700 064, India

Homi Bhabha National Institute, Training School Complex, Anushakti Nagar, Mumbai - 400094, India

E-mail: tilak@vecc.gov.in

The word "radioactive" was first coined by Marie Curie when she, along with her husband Pierre Curie, discovered the element Polonium. The nucleus $^{210}$Po is a testing ground for many theoretical and experimental aspects of nuclear structure as well as nuclear fission dynamics as it is a magic nucleus with neutron number N=126. At Variable Energy Cyclotron Centre, Kolkata the fission of Polonium nuclei is being studied in order to understand the survival of nuclear shell effects that is known to be the key for the stability of super heavy elements (SHE).

This year we are celebrating the 150th birth anniversary of Marie Skłodowska-Curie (1867-1934). She was the first woman to win a Nobel Prize and the first scientist to win two; one in the field of Physics (1903) and the other in Chemistry (1911). Polonium nuclei were discovered by Marie Curie and her husband Pierre Curie [1,2]. The substance that Curie couple extracted from pitchblende contained an unknown element, similar to bismuth in its analytical properties but much more active than uranium, the only active element known at that time. They proposed that the new element to be named Polonium in honour of the native land Poland of Marie Curie.

Polonium is also an interesting element to the nuclear physicists. One of its isotopes, $^{210}$Po has neutron number N=126 which is a magic number. The nuclei with the magic number of nucleons (neutron or proton) show extra stability as the outer nuclear shell is fully filled up. This is known as nuclear shell effect. A close analogy of magic nuclei is inert gases (Helium, Neon, Argon etc.) which are chemically inert because of their outermost electronic orbits are completely filled up. Usually with increase in atomic (proton) number, nuclei become unstable because of coulomb repulsion. However, it is predicted that due to the nuclear shell effect, elements having atomic number higher than 104, called super heavy elements (SHE), do exist. The current chart of the periodic table is being extended with newly discovered SHE [3]. Theoreticians predict that the super heavy element (SHE) with next neutron magic number N=184 and proton number Z=114/120/126 (they have different opinion about the proton numbers) will be a stable element. Efforts are on worldwide to reach this island of stability [4, 5].

However, to produce SHE one would require to find out the best combinations of nuclei (target and projectile) to be fused in the laboratory [6,7] . Also it is required to be ensured that the SHEs are produced at right excitation energies at which nuclear shell effects survives [8].

Since currently it is not possible to produce the element with N=184 (as that would require highly neutron rich targets and projectiles which are not available in nature), $^{210}$Po with magic number N=126 has been studied extensively to explore the survival of nuclear shell effects around the excitation energy around 40 MeV, beyond which one would expect the shell effects to vanish. Researchers of our Variable Energy Cyclotron Centre, Kolkata carried out several experiments recently [9, 10] to explore the existence of shell effects at fission saddle at moderate excitation energy for $^{210}$Po.

A few recent experimental and theoretical studies [11, 12, 13, 14], however, indicated dramatic ambiguity regarding the presence or absence of shell effects (correction) at the saddle point. Saddle point is the point of no return, so that a nucleus, if deformed beyond this point, will have no choice but to undergo fission, leading to two heavy fragments of nearly equal mass. Deviation of fission fragment angular anisotropy (which is the ratio of the fission yields along the beam direction and perpendicular to the beam direction) from theory was observed in the fission of $^{210}$Po, produced in nuclear reaction $^{12}$C+$^{198}$Pt, at excitation energy ~ 40-60 MeV. This was conjectured as an indirect evidence of shell effect at saddle due to neutron shell closure at N=126 [13].

During the fission process, many neutrons are also emitted, both before and after the scission. The neutrons emitted prior to fission are called the pre-scission neutrons. They are likely to carry signature of the dynamics of the process and therefore can be used as an independent experimental probe, similar to angular anisotropy. The pre-scission neutron multiplicity, measured in an experiment at the Inter University Accelerator Centre (IUAC), New Delhi, for $^{206,210}$Po. The result also indicated [12] the requirement of substantial shell correction not only in shell closed $^{210}$Po but also in $^{206}$Po. The analysis of both the angular anisotropy and pre-scission neutron data mentioned above, were carried out within the framework of the well-established statistical models. However, a recent dynamical calculations using stochastic Langevin equation, Schmitt et al [11] claimed that the angular anisotropy and neutron data, mentioned above, could well be explained with a purely macroscopic potential energy landscape without considering any shell effect at saddle point. Recent statistical model calculation [15] carried out by the scientists of Bhabha Atomic Research Centre for the nucleus $^{210}$Po populated in light and heavy ion induced reaction, however could describe the excitation functions without the requirement of shell correction at saddle, but required a huge fission delay to fit the pre-scission neutron multiplicity data in heavy ion induced reaction. This prevailing dramatic ambiguity necessitated an immediate evaluation of the problem through a new experimental observable.

The fission fragment mass distributions can be used as an experimental probe to look for the presence or absence of shell effects in nuclei [8]. Therefore mass distributions of the nuclei $^{210,206}$Po were studied [9] by the researchers at Variable Energy Cyclotron Centre. The nuclei were populated by bombarding $^{12}$C projectile from Tata Institute of Fundamental Research – Bhabha Atomic Research Centre pelletron machine on $^{198,194}$Pt targets. Large area multi-wire proportional counters (MWPC) [16], developed at Variable Energy Cyclotron Centre were used in the experiment to detect the fission fragments. Variation of the width of the fission fragment mass distributions with excitation energies were studied in the experiment.

The presence of shell effect would make the fission mass distributions either double humped or wider (compared to the theoretical estimate). Figure 1 show typical measured mass distributions at similar excitation energies for both the nuclei studied $^{210}$Po and $^{206}$Po. The mass distributions could be well fitted by single Gaussians as shown by the red line. The variation of standard deviation of the fitted mass distributions with excitation energy, shown in figure 2, was found to increase monotonically, consistent with theoretical estimate without incorporation of shell effect at the saddle point. This clearly indicated the absence of shell correction at the saddle point in both the Polonium isotopes up to the excitation energy 40 MeV. The experimental result [9] provides benchmark data to test the new fission dynamical models to study the effect of shell correction on the potential energy surface at fission saddle point.

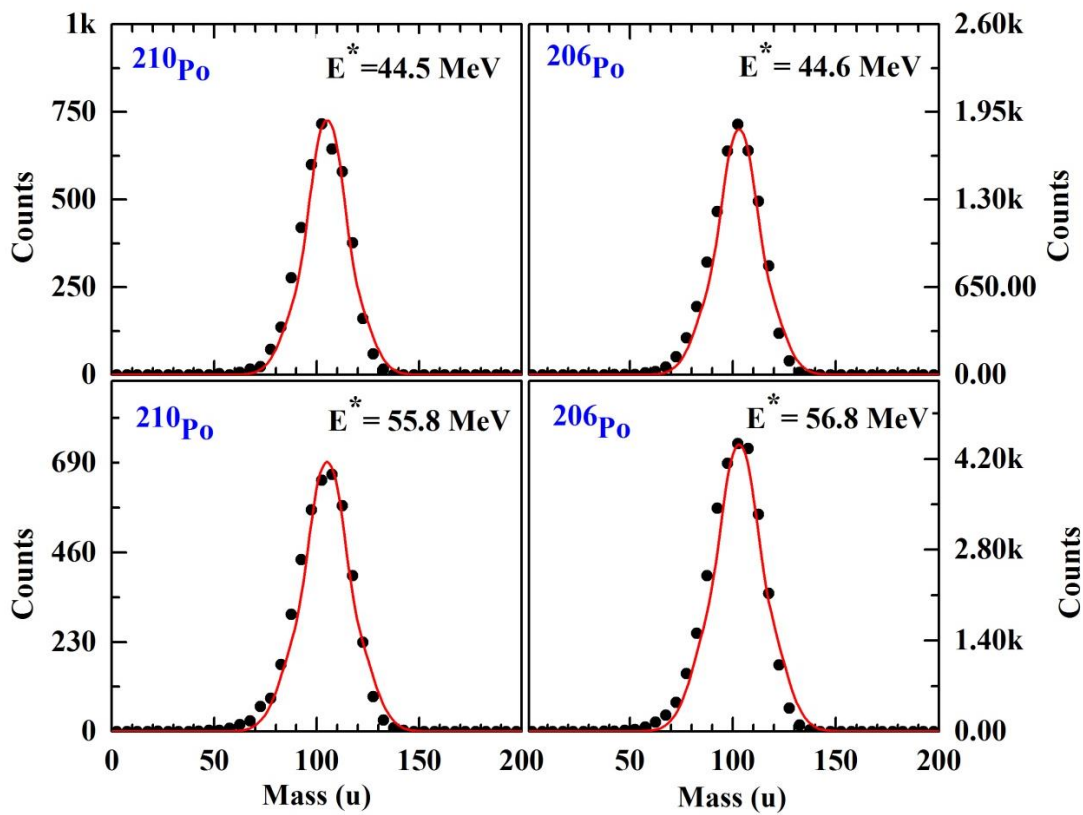

*Figure 1: Measured mass distributions of fission fragments at similar excitation energies for the two Polonium isotopes. At these excitation energies, the mass distributions are single peaked and could be fitted by a single Gaussian distributions as shown by solid (red online) lines.*

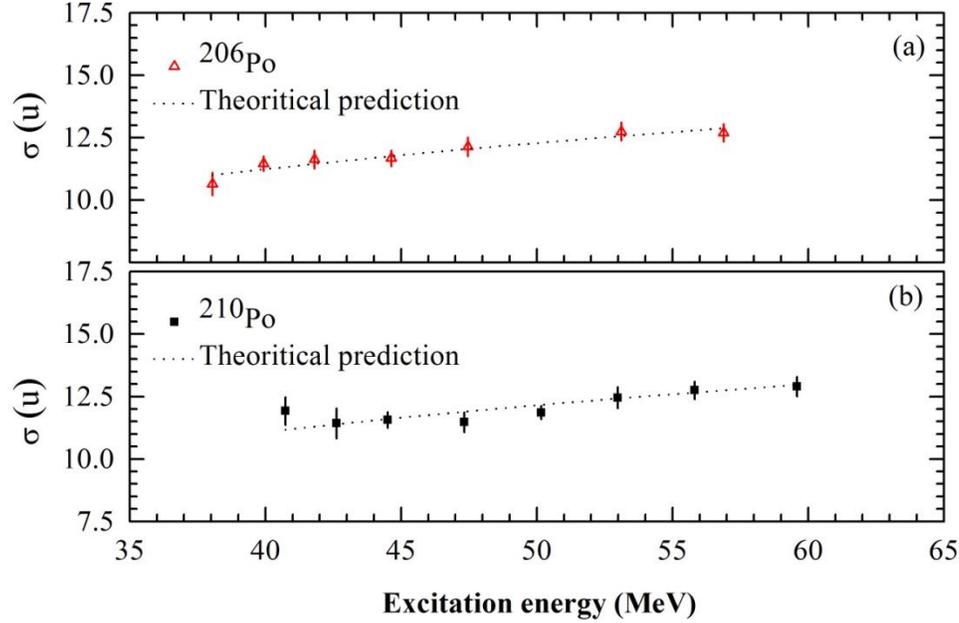

*Figure 2: Variation of the standard deviation of the fitted symmetric mass distribution with excitation energy. The calculated standard deviations from theory [9] are shown by dotted lines.*

Soon after the discovery [17] of nuclear fission, one of the greatest challenges was to explain the observation of asymmetric (double hump) mass distribution by Lise Meitner [18] while liquid drop model [19] predicts symmetric mass distribution. The asymmetry in mass distribution was understood by considering fragment shell properties [20]. It was conjectured that the 2nd peak (around mass number 132) of the mass distribution is due to the doubly magic shell configuration. Thus it was believed for years that shell effects of the fission fragments are responsible for this asymmetric mass distribution. However a recent experiment carried out at CERN ISOLDE facility shattered the notion.

In the pioneering experiment carried out CERN [21], fragment mass distribution of beta delayed fission of $^{180}$Hg was measured. The excitation energy of the nucleus was low (~ 10 MeV). At this excitation energy shell effects is expected to survive. One would also expect that the mass distribution of $^{180}$Hg be symmetric (peaking around mass number 90, with semi magic proton number Z=40 and neutron number N=50) if shell effects of the fission fragments would decide the fate of the mass distribution. However, the measured mass distribution was found to be asymmetric, peaking around mass number 80 and 100 (neither of which is near magic number), clearly contradicting the expectation.

Several theories came up to explain the puzzle of asymmetric mass distribution of $^{180}$Hg [22, 23]. These theories claim that apart from the shell effects of the fission fragments, the shell effects during transition from saddle to scission (known as fission valley) play the decisive role in deciding the fission fragment mass distribution. The macroscopic microscopic model (Brownian shape motion on five-dimensional (5D) potential energy surface) of Peter Moller and Jorgen Randrup could successfully explain the asymmetric mass distribution of $^{180}$Hg [22]. However, a good theory should have a predictive power of the mass distributions of several nuclei. Moller's theory predicted asymmetric mass distribution for $^{210}$Po nuclei at an excitation energy of 31.43 MeV, while at higher excitation energies mass distribution are predicted to be symmetric.

In order to test their theories, in a recent experiment conducted at the VECC K-130 cyclotron, alpha beam was bombarded on an enriched $^{206}$Pb target to populate the nuclei $^{210}$Po. At lower excitation energies fission cross sections are low (few mill barn). However, indigenously developed at VECC the large area multi wire proportional counters (MWPC) that provide large detection coverage helped us to successfully measure such a low cross section. The photograph of the experimental setup is shown in Figure 3. The masses of the fission fragments were calculated from the time of flight difference of the fragments and their positions (theta, phi) measured with the MWPCs.

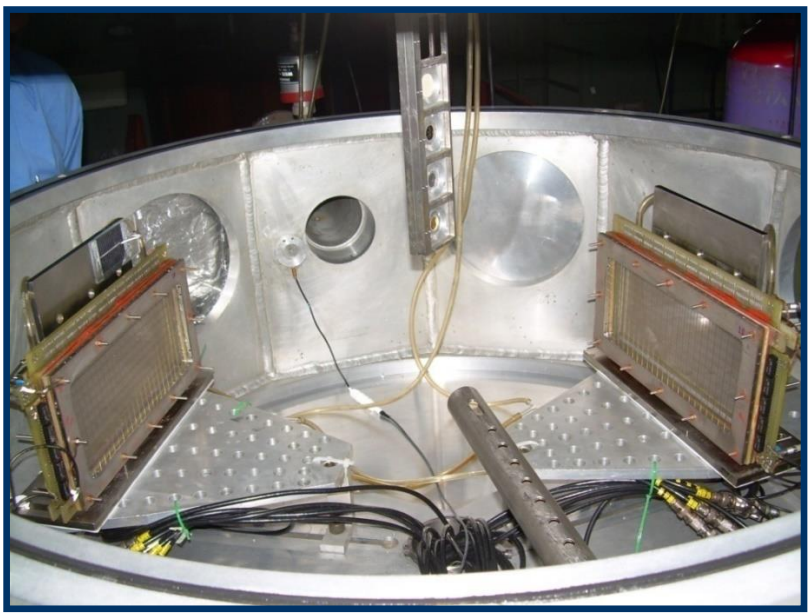

*Figure 3 : Photograph of the MWPC detectors mounted inside a scattering chamber for the detection of fission fragments. The detectors were indigenously developed at VECC.*

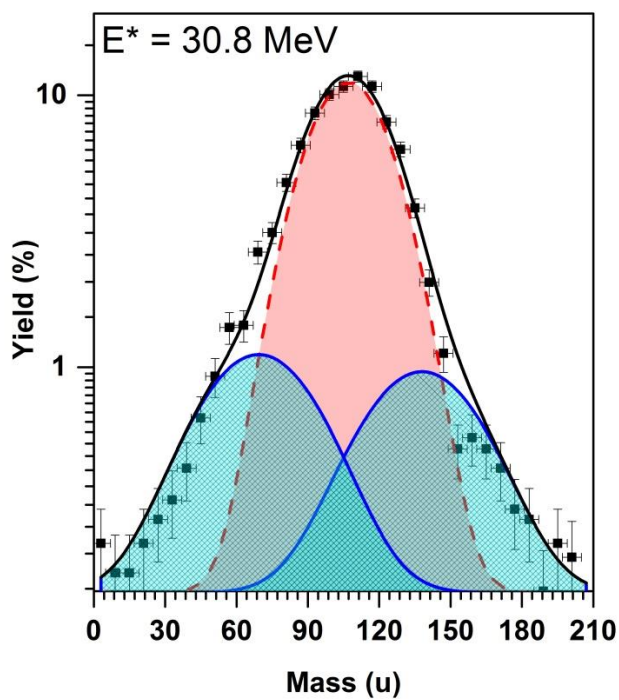

*Figure 4: Asymmetric mass distributions observed in the fission of $^{210}$Po nuclei at the excitation energy 30.8 MeV. This verifies a recent theoretical calculation [20] to predict the mass distribution of Polonium and other nuclei (e.g; $^{180}$Hg). The red dashed line indicates the*

*Gaussian fit corresponding to the symmetric fragments, while the blue solid lines show the asymmetric components. The overall fitting is shown by the solid black line.*

The measured fragment mass distributions [10] for the reaction $^4$He + $^{206}$Pb populating $^{210}$Po were found to be symmetric and the width of the mass distributions were found to increase monotonically with excitation energy above 36 MeV, indicating the absence of shell effect at the saddle. However, deviation from symmetric mass distributions at the lowest excitation energy (30.8 MeV) was indeed observed as shown in Figure 4. At this excitation energy, a weak asymmetric component (≈1% yield level) was shown to coexist along with the strong symmetric mass distribution. The heavier mass side of the asymmetric distribution was observed to peak around ≈132 (doubly magic $^{132}$Sn), a clear indication of the persistence of the shell effect of the nascent fragments. VECC experimental data [10] could reaffirm the theory [22].

It is clear from the above discussions that shell effect on the evolution of fission dynamics is still not fully understood. The Polonium case presented above showed a classic example of different facets of shell effects unfolded through the use of different experimental probes. The experimental fission research program at VECC to measure the fission fragment mass distributions of Polonium nuclei, provide strong support that the shell effects at fission valleys (i.e; during transition from saddle to scission) plays an important role. This indeed changes our notion that it's only the shell effects of the fission fragments that decide the fate of the mass distributions.

Polonium was discovered by Curie couple. In Marie Curie's 150$^{th}$ birth anniversary, we the nuclear physicists offer our tribute, for her important discovery of the element which still plays important role in the frontier area of nuclear physics research.

**Acknowledgements:**

The article is dedicated to Prof. Sailajananda Bhattacharya (former Head of Physics Group, VECC) on the eve of the completion of his tenure as a Raja Ramanna Research Fellow. Thanks to his encouragement, support and active participation that an experimental fission research program was initiated at Variable Energy Cyclotron Centre, Kolkata.

The author acknowledge to all his collaborators and accelerator staffs (VECC, Kolkata and BARC-TIFR Pelletron facility, Mumbai, IUAC, New Delhi) who have contributed in this experimental research program. The author is thankful for the support from all his students: Abhirup Chaudhuri, Trinath Banik, Arnab Ghosh and colleagues: C. Bhattacharya, Arijit Sen, K. Banerjee, S. Kundu, T.K. Rana, J. K. Meena, G. Mukherjee, P. Roy, R. Pandey, S. Manna, J. K. Sahoo, A. Saha, R. Saha Mondal.